\documentclass[12pt]{article}

\usepackage{pslatex}
\usepackage{float}
\usepackage{amsmath}
\usepackage{color}
\usepackage{graphicx}
\usepackage{amssymb}
\usepackage{subfigure}
\makeatother
\begin{document}

\begin{flushright}
{\bf  
IFJPAN-V-2005-8\\   CERN-PH-TH/2005-139 \\
} 
\end{flushright}
\begin{center}
{
     \Large\bf
PHOTOS as a pocket parton shower:  \\
flexibility tests for the algorithm}
\end{center}

\vspace*{1mm}

\begin{center}
 {\em Partial documentation of  work performed for  the HERA4LHC workshop }
\end{center}

\vspace*{1mm}


\begin{center}
 
{\bf Piotr Golonka}\; and \; {\bf Zbigniew Was}
 \vspace{5mm}

{\em CERN, 1211 Geneva 23, Switzerland} \\ and \\
{\em Institute of Nuclear Physics, P.A.S., ul. Radzikowskiego 152, 31-342 Krak\'ow, Poland}

\end{center}
\vfill

Recently version 2.14 of the {\tt PHOTOS} Monte Carlo algorithm, written for
bremsstrahlung generation
in decays became available.  In ref. \cite{Golonka:2005pn}, detailed
instructions on  how to use the program are given.
With respect to older versions \cite{Barberio:1990ms,Barberio:1994qi} 
of  {\tt PHOTOS}, it now features: 
 improved implementation of QED interference and multiple-photon 
radiation. The numerical stability of the code was significantly improved as well.
Thanks to these changes, {\tt PHOTOS} generates bremsstrahlung corrections 
in $Z$ and $W$ decays with a precision of 0.1\%. This 
precision was established in \cite{tauolaphotos} 
with the help of a multitude of distributions and of a specially 
designed test (SDP).

In this note we will not repeat a discussion of the design properties,
but we will recall the main tests that document robustness and flexibility 
of the {\tt PHOTOS} design. This aspect may be of broader use and may find extensions
in future applications, also outside the simple case of purely QED bremsstrahlung 
in decays.

We begin with an informal  presentation of  the 
components of the  {\tt PHOTOS} algorithm using operator language.
The consecutive approximations used in the construction of 
the crude distribution for photon generation, and the 
correcting weights used to construct the physically complete distributions
are listed, but will not be defined in detail, because of limited space
allowed for the workshop contributions. Instead, we present the variations of the algorithm. 
Comparisons between different options of the algorithm provide an important
class of technical tests, and also helps  to explore 
 limits of the universality of the  {\tt PHOTOS} solution.
The results of some of these tests will be listed later in the contribution
(for the remaining ones and details
we address the reader to refs. \cite{Golonka:2005pn,tauolaphotos}).
In the comparisons we use the SDP universal test based on  {\tt MC-TESTER} \cite{Golonka:2002rz} as  in ref. \cite{Golonka:2005pn}. 
We must  skip the repetition of its definition here as well.

The starting point for the development of  {\tt PHOTOS} was the observation
that, at first order, the bremsstrahlung corrections in the
$Z \to \mu^{+} \mu^{-}$ process
can be written as a convolution of the Born-level distribution with 
the single-photon emission kernels for the emission from  
$\mu^+$ and $\mu^-$. 

\vspace*{1mm}
\bigskip
\footnoterule
\noindent
{\footnotesize \noindent
Supported in part  by the EU grant MTKD-CT-2004-510126, 
in partnership with the CERN Physics Department,
and the Polish State Committee for Scientific Research 
(KBN) grant 2 P03B 091 27 for the years 2004--2006.
}
\newpage
\normalsize


The formulae for the emission kernels are 3-dimensional
and can be parametrized using the angles and the invariant mass, 
which are the same variables as those used in the
parametrization of the three-body phase space
(the kernels use only a subset of the complete set of
phase-space parametrization variables).
The remaining two angular variables in the kernels
can be identified as the angles defining the
orientation of the $\mu^+$ and/or $\mu^-$ directions (for a detailed definition,
see e.g. \cite{Barberio:1990ms}).

The principle of the action of the single-photon algorithm working
on $n$-body decay is to replace 
a point in the $n$-body phase space $\Omega_2$, with 
either the point in the original $\Omega_2$,
or the point in the $(n+1)$-body phase space $\Omega_3$ (with 
generated photon). The overall normalization of the decay rate has to change 
as well and, for example, in the case of $Z\to \mu^+\mu^-$ 
it needs to be multiplied by a factor of 
$1+ \frac{3}{4}\frac{\alpha}{\pi}$.

Subsequent steps of
the  {\tt PHOTOS} algorithm are described in terms  of the evolution operators.
Let us stress the relations of these operators to the  matrix elements and 
phase-space parametrizations.
We will present the decomposition of the operators in 
the top--down order, starting with the definition of 
$R_\alpha$, the operator describing the complete  {\tt PHOTOS }
algorithm for single emission (which at least 
in the case of $Z$ and leptonic $\tau$ decays originates from
field theory calculations without any approximation).
 Then, we will gradually decompose the operators
(they differ from decay channel to decay channel)
 so that we will end up with the single well-defined,
elementary operator for the emission from a single charged particle in the final state.
 By aggregation of these elementary
operators,  the $R_\alpha$ may be reconstructed for any decay channel.
Let us point out that the expression of theoretical calculations
in the form of operators is particularly suitable 
in computer programs  implementation.

We cannot present here
a separate discussion of the factorization properties, 
in particular to define/optimize the way the iteration
of $R$'s is performed in  {\tt PHOTOS}.
Not only the first-order calculations are needed, but also
higher-order ones, including mixed virtual--real corrections.
For practical reasons, the $R_\alpha$ operator
needs to be regularized with the minimum energy 
for the explicitly generated photons:
the part of the real-photon phase space, under threshold, 
is integrated, and the resulting factor is summed with
the virtual correction.

\centerline{$\bullet$ 1} 

Let us define the five steps in $R_\alpha$ separation.
In the first one, the 
$R_\alpha$  is replaced by (we use two-body decay  as an example)
$R_\alpha =  R_I(R_S(\mu^+) + R_S(\mu^-)) $,
where $R_I$ is a generalized interference operator,
and $R_S$ is a generalized operator responsible for photon
generation from a single, charged decay-product. 
Here we understand the generalized interference operator as  
shifting between different kinematical configurations while respecting energy--momentum
conservation, thus also overall normalization of the distribution under construction. 

There is a freedom of choice in the  separation of $R_\alpha$ into $R_I$ and $R_S$.
The $R_S$ operator acts on the points from the $\Omega_2$ phase space,
and the results of its action belong either to $\Omega_2$ 
or to $\Omega_3$. The domain of the $R_I$ operator has to be
$\Omega_2$ +  $\Omega_3$, and the results are 
in either $\Omega_2$ or  $\Omega_3$. 
In our solution we required that $R_I$ acts 
as a unit operator on the $\Omega_2$-part of its domain
and, with some probability, returns the points from  
$\Omega_3$ back to the original points in $\Omega_2$,
thus reverting the action of the $R_S$. 

Let us stress that in practical applications,
to ease the extension
of the algorithm to ``any'' decay mode,
 we
used in {\tt  PHOTOS} a simplification for $R_I$ . 
Obviously, the exact representation of the first-order result 
would require $R_I $ to be decay-channel-dependent.
Instead, we used an approximation that ensures the  proper 
behaviour of the photon distribution in the soft limit.
Certain deficiencies at the hard-photon limit of the phase space
appear as a consequence, and 
are the subject of studies that need to be performed 
individually for every decay channel of interest. 
The comparisons with matrix-element formulae, as 
in \cite{Nanava:2003cg}, or 
experimental data, have to be performed 
for the sake of precision; they may result 
in dedicated weights to be incorporated into  {\tt PHOTOS}.
In principle, there is no problem to install a particular decay-channel
matrix element, but there
has not been much need for this yet. So far, the precision of the  {\tt PHOTOS}
algorithm could always be raised to a satisfactory level by implementing
some excluded parts of formulae,  being  the case of $W$ decay 
\cite{Nanava:2003cg} an exception.

The density generated by the $R_S$ operator is typically twice that of real photons
all over the phase space; it can also overpopulate only those regions
of phase space where it is necessary for $R_I$.
The excess of these photons is then reduced by Monte Carlo with the action of 
$R_I$.

\centerline{$\bullet$ 2} 

In the next step of the algorithm construction, we have separated 
$R_S=R_B R_A$, where $R_B$ was responsible for the implementation of
the spin-dependent part of the emission, 
and the $R_A$ part was independent of the spin of the 
emitting final-state particle.
Note that this step of the algorithm can be performed at the earlier 
stage of generation as well, that is before the full angular construction of the event. 
$R_B$ is again, as $R_I$, of the generalized interference type; it moves the hard bremsstrahlung 
events in excess back to the original no-bremsstrahlung ones. $R_B$ operates on the internal variables of 
{\tt PHOTOS} rather than on the fully constructed events.

\centerline{$\bullet$ 3}
 
The definition of the $R_I$, $R_B$, $R_A$ operators was initially based 
on the inspection of the first-order matrix elements for the two-body decays.
 In the general solution for $R_A$, the process of multiple-body decay 
of particle $X$ is temporarily replaced  by the two-body decay  $X \to ch Y$, 
in which particle $X$ decays to the charged particle $ch$, which  ``emits'' 
the photon,
and the ``spectator system'' $Y$. The action of the operator is repeated
for each charged decay product: the subsequent charged particle takes the role
of the photon emitter $ch$; all the others, including the photons generated in
the previous steps, become a part of  the spectator system $Y$. The independence
of the emissions from each charged product then has to be ensured.
This organization works well and can be understood 
with the help of the exact parametrization of multibody phase space. 
It is helpful for iteration in  multiple-photon emission.
It also helps to implement some genuine second-order matrix elements.
This conclusion can be drawn from an inspection of the second-order 
matrix elements, as in \cite{Was:2004ig}.

\centerline{$\bullet$ 4} 

In the next step, we decompose the $R_A$ operator, splitting it in 
two parts: $R_A =R_a R_x$.
The $R_x$ operator generates the energy of the (to be generated) photon,
and $R_a$ generates its explicit kinematical configuration.

The $R_x$ operator acts on points from the $\Omega_2$ phase space,
and generates a single real number $x$;  
the $R_a$ operator transforms this point from $\Omega_2$ and the number $x$ 
to a point in $\Omega_3$, or leaves the original point in $\Omega_2$. 
Note that again, as $R_I$, the $R_a$ operator has to be  unitary and has to conserve 
energy--momentum. 
The $R_x$ operator does not fulfill these criteria. 

An analogy between $R_x$ and the  kernel for structure-function evolution should be 
mentioned.
However, there are notable differences:
the $x$ variable is associated more with the ratio of the invariant
mass of decay products of $X$, photon excluded, and the mass of $X$,  than with
the fraction of energy taken away by the photons from the
outgoing charged product $ch$. Also, $R_x$ can be simplified by moving its
parts to 
$R_a$, $R_S$ or even $R_I$. Note that in  $R_x$ the contributions of radiation from all 
charged final states are summed.

\centerline{$\bullet$ 5} 

The $R_x$ operator is iterated, in  the solutions for double, 
triple, and  quartic photon emission. 
The iterated $R_x$ can also be 
shifted and grouped at  the beginning of the generation,
ignoring phase-space constraints. The iterated $R_x$ takes a form similar to a formal
solution for structure-function evolution, but with exceptionally simple kernels. 
The phase-space 
constraints can then be introduced later, with the action of the $R_a$ operators. 
Because of this, the iteration of $R_x$  can go up
even to  infinite order. The algorithm is
then organized in two steps. At first, a crude distribution for the number of photon
candidates is generated; then, their energies are defined.
At this stage we can perform a further separation: $R_x= R_f R_0$,
where the $R_0$ operator determines whether a photon candidate has to be
generated
at all, and $R_f$ defines the fraction of its energy 
(still without energy--momentum-conservation constraint). From the iteration of  $R_0$,
we obtain a Poissonian distribution, but any other analytically solvable distribution would be equally good.

The overall factor, such as  $1 +\frac{3}{4} \frac{\alpha}{\pi}$
in $Z$ leptonic partial width, 
does not need to be lost.
It needs to find its way to the $R_0$,
and affects the total rate of the process.
For the case of the FSR, discussed here, we can skip this point;
however, it may be important for generalizations. 

\centerline{ --- } 

The input data for the algorithm are taken from the event
record, the kinematical configurations of all particles, and the 
mother--daughter relations between particles in the decay
process (which could be a part of the decay cascade) should be
available in a coherent way.

This wraps up, a basic, presentation of the steps performed by
the  {\tt PHOTOS} algorithm. For  more details see \cite{Golonka:2005pn,PhDGolonka}.

\vskip 2 mm
%
\centerline{\label{sec:tests} \bf Tests performed on the algorithm:}
%

\begin{enumerate}
\item The comparison of  {\tt PHOTOS} running in the quartic-photon emission mode
and the exponentiated mode for the leptonic $Z$ and $W$ decays may be found on our 
 web page which documents the results of the tests \cite{tauolaphotos}. 
The agreement in branching ratios and shapes of the distributions 
is better than 0.07\%
for all the cases that were tested. It can be concluded that changing the relative
order for the iterated $R_0$ and the rest of $R_\alpha$ operators does not
 lead to 
significant differences.
This test, if understood as a technical test,  is slightly biased  
by the uncontrolled higher-than-fourth-order
terms which 
are missing in the quartic-emission option of  {\tt PHOTOS}.
Also, the technical bias, due to the minimal photon energy in generation,
 present in the fixed-order
options of  {\tt PHOTOS} may contribute to the residual difference.

\item The comparison of    {\tt PHOTOS} with different options for the relative separation 
between $R_I$ and $R_S$. 
The tests performed for the fixed-order and exponentiated modes indicated
that the differences in results produced by the two variants of the
algorithm are below the level of statistical error for the runs of
$10^8$ events. In the code these two options are marked respectively as
{\tt VARIANT-A}  and {\tt VARIANT-B}.

\item The comparisons of  {\tt PHOTOS} with different algorithms for the 
implementation of the $R_I$ operator. In  {\tt PHOTOS} up to version 2.12,
the calculations were performed using internal variables in the
angular parametrization. This algorithm was limited to the cases
of decays of a neutral particle into two charged particles.
In later versions, the calculations are performed using the 
4-momenta  of particles, hence for any decay mode.
The tests performed for leptonic $Z$ decays indicated that the
differences are below the statistical error of the runs of $10^8$ events.

\item Comparison of  {\tt PHOTOS} with different options for the relative separation 
between $R_0$ and  $R_x$, consisting of an increase in the
crude probability of hard emission at $R_0$.
The tests performed for the exponentiated mode of  {\tt PHOTOS} indicated that
the differences are below the statistical error of the runs 
of up to $10^8$ events.

\item The remaining tests,
including new tests for the effects of the interference weights
in cascade decays, are more about the physics content of the program
 than on the technical or algorithmic aspects. They
are presented in ref. \cite{Golonka:2005pn} and the results are 
collected on the web page \cite{tauolaphotos}. 
\end{enumerate}

Multiple options for  {\tt PHOTOS} running and technical compatibility of results 
even for $10^8$ event samples generated in a short CPU cycle time are encouraging. 
They indicate the potential for algorithm extensions. Note that  {\tt PHOTOS} was found 
to work  for decays of up to 10 charged particles in the final state.

{\bf Acknowledgements:} The authors are indebted to members of BELLE, BaBar, NA48, KTeV,
ATLAS, CMS, D0, CDF collaborations for useful comments and suggestions.


\providecommand{\href}[2]{#2}\end{document}